**Transboundary Secondary Organic Aerosol in the Urban Air of Fukuoka, Japan**

Satoshi Irei,[1†*] Akinori Takami,[1] Keiichiro Hara,[2] and Masahiko Hayashi[2]
[1]National Institute for Environmental Studies, 16-2 Onogawa, Tsukuba, Ibaraki 305-8506, Japan
[2]Department of Earth System Science, Faculty of Science, Fukuoka University, 8-19-1 Nanakuma, Jonan-ku, Fukuoka 814-0180, Japan
Present address: Department of Chemistry, Biology, and Marine Science, University of the Ryukyus, 1 Senbaru, Nishihara, Okinawa 903-0213, Japan
[*]Corresponding author: Satoshi Irei (satoshi.irei@gmail.com)

**Abstract:** To understand the influence of transboundary secondary organic aerosol (SOA) in the urban air of Fukuoka, we conducted simultaneous field studies in December 2010 and March 2012 on Fukue Island and in the city of Fukuoka, representing a rural and an urban sites of western Japan. During the studies, organic aerosol in $PM_{1.0}$ was analyzed by aerosol mass spectrometers. Independently, airborne total suspended particulate matter (TSP) was collected on filters, and low-volatile water soluble organic carbon (LV-WSOC) extracted from the TSP samples was analyzed by an elemental analyzer coupled with a high precision isotope ratio mass spectrometer for its concentration and stable carbon isotope ratio ($\delta^{13}C$). Plots of LV-WSOC concentrations versus *m/z* 44 concentrations for the study in December 2010 showed high correlations ($r^2 > 0.85$), but different slopes between the two sites (0.58 μg $\mu gC^{-1}$ at Fukuoka and 0.22 μg $\mu gC^{-1}$ at Fukue), indicating additional contribution of *m/z* 44 from local sources. Because of this complex composition, evaluation of transboundary SOA was unsuccessful in this case study. In contrast, comparison of LV-WSOC concentrations with the *m/z* 44 concentrations during the study in March 2012 showed high correlations ($r^2 > 0.91$) with similar slopes (~ 0.3 μg $\mu gC^{-1}$) at both sites, indicating predominance of transboundary LV-WSOC. Furthermore, the plot of $\delta^{13}C$ of LV-WSOC as a function of *m/z* 44 fraction showed a systematic increasing trend, indicating SOA. These findings suggest that the observed LV-WSOC and *m/z* 44 in the Fukuoka air were predominated by transboundary SOA in this case study.

## 1. Introduction

Recent severe air pollution in eastern China has been one of serious environmental issues for neighboring countries in the north-east Asia because the prevailing wind during fall, winter, and spring seasons in this region often carries pollution across the country boundary. It is suspected that air quality in Japan is influenced by this transboundary air pollution. Fukuoka is the largest city in the Kyushu region of Japan, which is the westernmost region closest to the Chinese continent. Approximately 1.5 million people reside in the city of Fukuoka, and thus with respect to public health, a quantitative understanding of transboundary pollution in the local air is urgent. When evaluating air quality, assessments should consider not only primary pollutants, which are emitted directly from emission sources, but also secondary pollutants, which are formed in the air by oxidation of primary pollutants, such as ozone, peroxyacetyl nitrate, nitro polycyclic aromatic hydrocarbons, and secondary organic aerosol (SOA). The estimation of atmospheric SOA is important due to its potential association with cloud formation (Anderson *et al.*, 2003) and adverse health effects (Dockery *et al.*, 1993). Nevertheless of such importance, its quantitative understanding has been very limited because most of oxidation products are ubiquitous and no unique molecular marker has been identified to date. To the best of our knowledge, positive matrix factorization (PMF) analysis of organic aerosol mass spectra obtained by aerosol mass spectrometry (AMS) measurements may only be the method currently used for evaluation of atmospheric SOA (Zhang *et al.*, 2005: Ng *et al.*, 2010). Recently, field studies demonstrated that stable carbon isotope ratio ($\delta^{13}C$) measurements of particulate low-volatile water soluble organic carbon (LV-WSOC) at rural sites in the Kyushu and Okinawa regions captured evidence of SOA (Irei et al., 2014). Due to no large emission source of anthropogenic SOA precursors near the sites and the evidence of air mass transport from the Asian continent during the study period, they concluded that the observed LV-WSOC was strongly associated with SOA converted from transboundary volatile organic substances. Because these observations were made in the westernmost Japan, the urban air of Fukuoka would also contain LV-WSOC associated with the transboundary SOA and may contain LV-WSOC from local origin.

Our objective in this study is to better understand the composition of LV-WSOC in the urban air of Fukuoka using techniques of $\delta^{13}C$ and AMS measurements. We conducted simultaneous field measurements in March 2012 and December 2010 at Fukuoka and Fukue, representing urban and rural sites of the northern Kyusyu region, respectively. Airborne particulate matter was collected noon to noon on quartz fiber filters, and the collected samples were analyzed for the carbon concentration and $\delta^{13}C$ of LV-WSOC. During the studies, we also analyzed an organic component of submicron aerosols using AMSs. The data obtained at the two sites with the two measurement techniques were then compared to evaluate origins of LV-WSOC.

## 2. Materials and Methods

We conducted simultaneous field studies at the Fukue atmospheric monitoring station (32.8°N, 128.7°E) and at Fukuoka University (33.6°N, 130.4°E) (Figure 1) during two periods: from 6 to 17 December, 2010 and from 7 to 18 March, 2012. The measurements and results of the field studies in 2010 at both sites and those in 2012 at Fukue have been presented previously (Irei *et al.*, 2014; Irei *et al.* 2015). Therefore, only the experiments during the study at Fukuoka in March 2012 are described here.

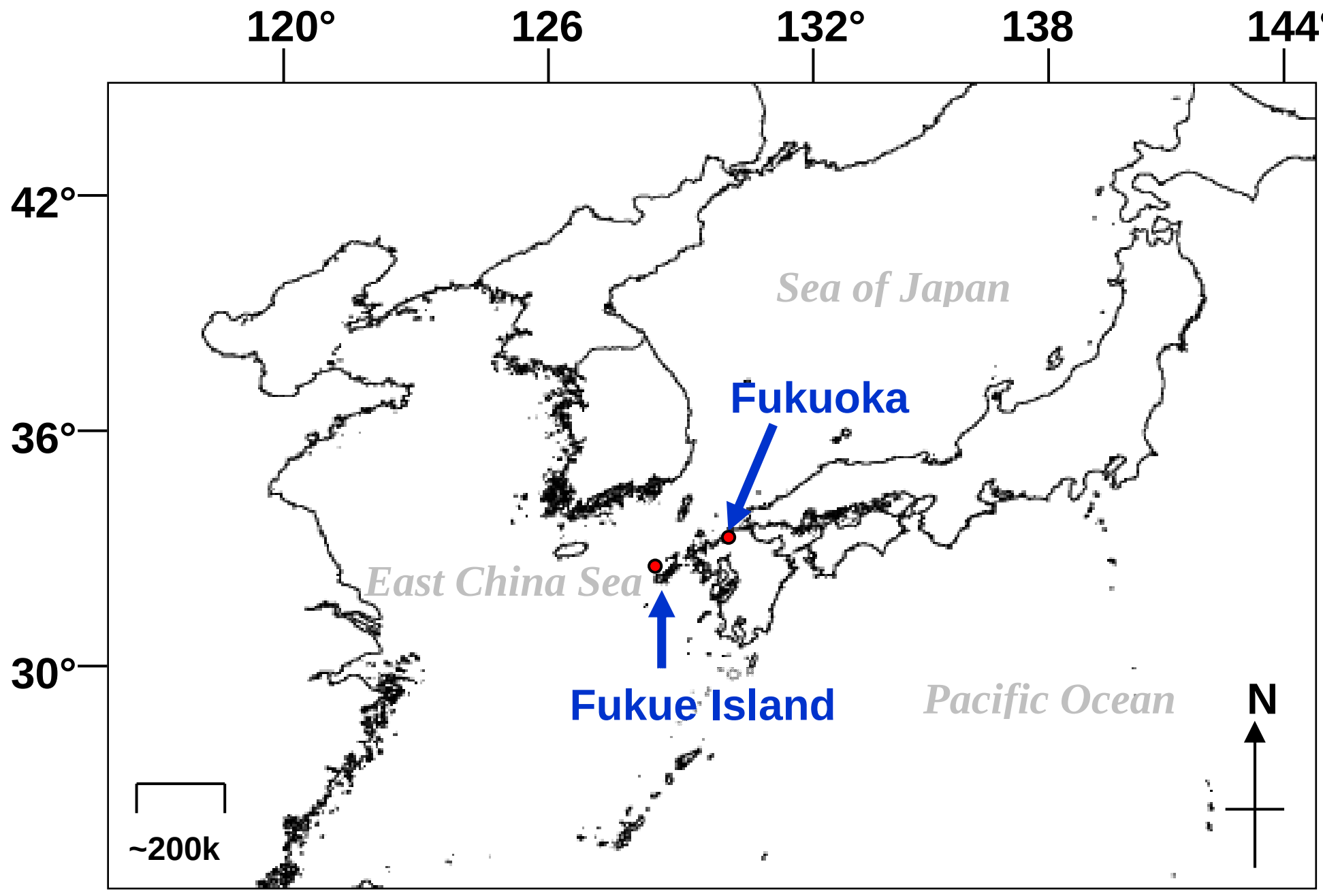

**Figure 1. Map showing the locations of measurement sites.**

Total suspended particulate (TSP) was collected every 24 h, noon to noon, on pre-baked (773 K for 12 h) quartz fiber filters (8 × 10 inch Tissuquartz, Pall Corp., New York City, NY, USA) at the rooftop of the Earth System Science Bldg. (~ 15 m above the ground) using a high-volume air sampler (HV-1000: Sibata Corp., Saitama, Japan). This building is surrounded by other buildings on campus. The sampling flow rate used was approximately 1 $m^3$ $min^{-1}$. We also collected one field blank and two field blanks were taken at Fukuoka and Fukue during the study in March 2012, respectively. We used a quarter of each filter sample for the LV-WSOC analysis. The details of the preparation procedure for LV-WSOC extracts have been described elsewhere (Irei *et al.*, 2014). Briefly, LV-WSOC was extracted with ultra-pure water (Wako Pure Chemical

Industries, Ltd., Osaka, Japan) via sonication. The extract was filtered (PURADISC 25 TF, Whatman Japan K.K., Tokyo, Japan) and then concentrated to ~0.1 mL using a rotary evaporator (R-205 and B-490: Nihon Büchi K.K., Tokyo, Japan) and a gentle flow of 99.99995% pure nitrogen (Tomoe Shokai, Tokyo, Japan). The final volume of the concentrated extract was determined by weighing the extract, assuming unit density. An aliquot of the concentrated extract was pipetted into a 0.15-mL tin cup for elemental analysis (Ludi Swiss AG, Flawil, Switzerland) and then completely dried under a gentle flow of pure nitrogen. Finally, a drop of 0.01 M hydrochloric acid (Wako Pure Chemical Industries, Japan) was spiked into the dried sample for the removal of carbonate, and the sample was re-dried. Extracted samples prepared in this manner were analyzed by an elemental analyzer (Flash 2000: Thermo Scientific, Waltham, MA, USA) coupled with an open-split interface (Conflo IV: Thermo Scientific), followed by an isotope ratio mass spectrometer (Delta V Advantage: Thermo Scientific) or EA-IRMS to determine the carbon mass and $\delta^{13}$C value. Here, $\delta^{13}$C is defined as follows:

$$\delta^{13}\mathrm{C} = \left[ \frac{(\frac{^{13}\mathrm{C}}{^{12}\mathrm{C}})_{\mathrm{sample}}}{(\frac{^{13}\mathrm{C}}{^{12}\mathrm{C}})_{\mathrm{reference}}} - 1 \right],$$

where $(^{13}\mathrm{C}/^{12}\mathrm{C})_{\mathrm{sample}}$ and $(^{13}\mathrm{C}/^{12}\mathrm{C})_{\mathrm{reference}}$ are the $^{13}$C/$^{12}$C atomic ratios for the sample and the reference (Vienna Pee Dee Belemnite), respectively. The method for determination of the LV-WSOC concentration and $\delta^{13}$C have been thoroughly evaluated previously (Irei *et al.*, 2014) by conducting standard spiked tests on sampling filters using the international (IAEA-C6 sucrose, the reference $\delta^{13}$C ± standard deviation or SD = -10.8 ± 0.5‰) and the laboratory standards (oxalic acid, the reference $\delta^{13}$C ± SD = -28.3 ± 0.2‰). Typical measurement reproducibilities (*i.e.*, standard deviations or SDs) for carbon concentration and $\delta^{13}$C are 11% and 0.3 ‰, respectively, but some measurement biases for those have been observed up to -23% and +0.4 ‰, respectively. Obtained results were not corrected for these possible biases.

We also used a quadrupole aerosol mass spectrometer (AMS, Aerodyne Research Inc., Billerica, MA, USA) to measure the chemical composition of fine aerosol (~$PM_{1.0}$). The details of these instruments and the theory for determining the concentration of chemical species have been described elsewhere (Jayne *et al.*, 2000; Allan *et al.*, 2004). Briefly, the AMS was set to output background-subtracted mean concentrations of chemical species over 10 minutes measurements. The scanning range of *m*/*z* was from 1 to 300, and the AMS flash vaporizer was set to 873 K. The instrument was calibrated with 300–350-nm dried ammonium nitrate particles at the beginning of the study period for quantitative analysis. Ionization efficiency determined was $6.2 \times 10^{-7}$ counts molecule$^{-1}$. Collection efficiency used for the concentration calculation was one, which was determined by comparing the sulfate concentrations measured by AMS with the non-sea salt sulfate concentrations determined by the filter sample analysis. Although AMS is capable to measure non-refractory particulate inorganic ions, such as ammonium, sulfate, and nitrate, and organic aerosol (OA), we focused on results of OA concentrations and *m/z* 44 signals in the OA mass spectra ($CO_2^+$ fragment ions originating from carboxylate of OA) in this paper. Based on the measurement results for filtered air, typical measurement reproducibilities for *m/z* 44 and OA concentrations are approximately 0.02 and 0.2 μg m$^{-3}$, respectively. The sampling for the AMS measurements was made from a room window at the 4th floor of the Earth System Science Bldg. (~ 10 m above the ground). Approximately 4 m × 0.5 in. o.d. stainless steel tubing (GL Science, Japan) was used for the sampling line, and a $PM_{2.5}$ separator (URG 2000-30ED, URG Corp., Chapel Hill, NC, U.S.A.) was attached to the inlet to cut off particles larger than 2.5 μm aerodynamic diameter.

Meteorological elements, such as precipitation, were also measured with a weather transmitter (WXT 520, VAISALA, Helsinki, Finland). Signals from the transmitter were recorded every 10 min using a data logger (NR-1000, KEYENCE, Osaka, Japan).

## 3. Results and discussion

### 3.1. Precipitation

During the study in March 2012, the weather transmitter at Fukuoka detected apparent precipitation in three periods; March 10th (3:30 ~ 5:20), 16th (10:50 ~ 21:00), and 18th (3:00 ~ 12:10) (Figure 2a). During this study precipitation was observed at Fukue in two periods; March 16th (8:10 ~ 22:10) and 17th through 18th (23:40 ~ 11:20) (Figure 2a). During the study in December 2010 at Fukuoka, apparent precipitation was observed in five periods; December 7th (9:30 ~ 22:00), 8th (4:30 ~ 8:40), 8th through 9th (18:30 ~ 2:00), 12th through 14th (22:30 ~ 11:20), and 17th (13:40 ~ 20:50) (Figure 2b). Precipitation during this study at Fukue was in three periods; December 8th through 9th (16:10 ~ 2:00), 12th through 14th (19:30 ~ 3:00), and 17th (3:00 ~ 16:40) (Figure 2b). Comparison of these time series plots exhibited that with respect to the timing of occurrence of precipitation events at both of the measurement sites roughly coincided, indicating the similar regional weather condition.

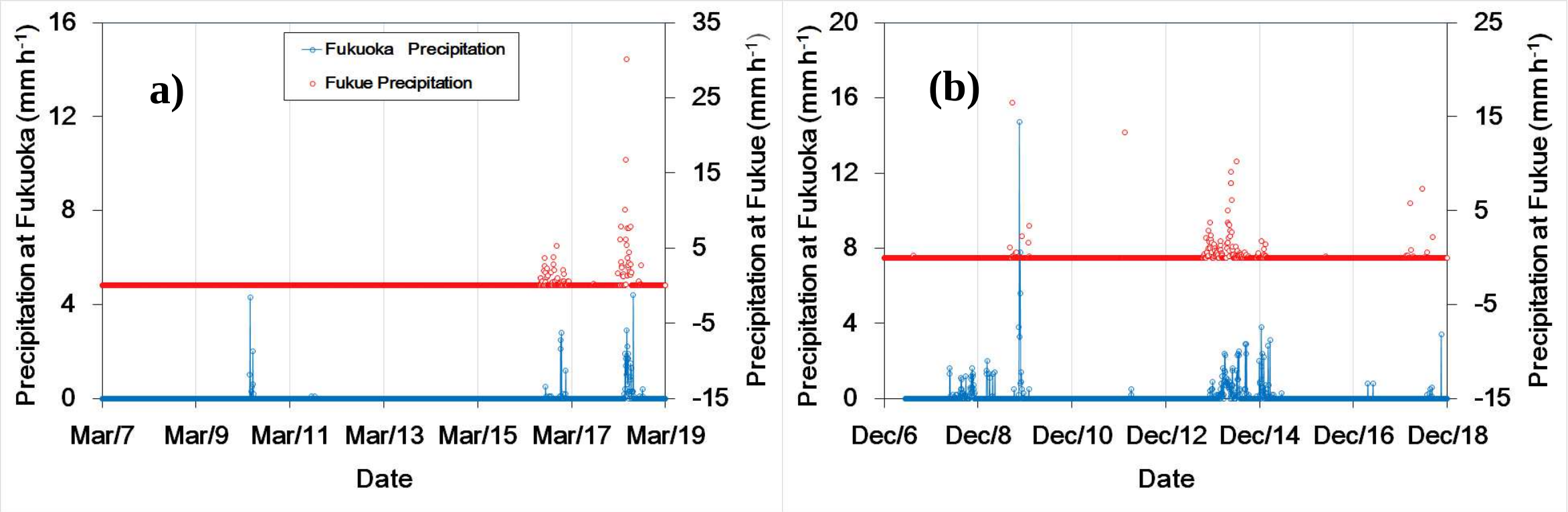

**Figure 2. Time series plot of precipitation observed at Fukuoka and Fukue during the field study in (a) March 2012 and (b) December 2010.** Time resolution in this plot is 10 min.

## 3.2. Concentration and $\delta^{13}C$

Results of AMS measurements showed that during the study in March 2012 the OA and *m/z* 44 concentrations at both sites were overall in similar magnitude, regardless of possible influence from local emissions at Fukuoka (Table 1). Except for the period between March 14 and 17, their overall time series variations at Fukuoka were also similar to those at Fukue (Figures 3a and 3b), suggesting that there were some common factors determining the OA and *m/z* 44 concentrations during the study period at both sites, such as transboundary air pollution and precipitation. Compared to the time series of precipitation observed during the study discussed earlier, the different variations in the concentration plots between the two sites during the period of March 14-17 are likely explained by the different time period of wet deposition at each site. Comparison between the *m/z* 44 and OA concentrations demonstrated similar variations at each site through the whole study period of March 2012, even though the magnitudes of their variations were different between the two sites (Figure 3a and 3b). The constant magnitude of OA and *m/z* 44 variations at Fukue indicates the constant ratios of *m/z* 44 to OA concentrations ($f_{44}$), implying the simple composition of OA, probably made of carboxylate-related substance(s). Meanwhile, the different magnitude of their variations at Fukuoka is explained by the irregular contribution of OA from the local man-made emissions. This is indicated by the observations of sharp rises and falls in the OA concentration plot.

**Table 1. Summary of AMS and LV-WSOC measurements.**

| | Number of available data | Mean | SD | Median | Upper quartile | Lower quartile | Max | Min |
|---|---|---|---|---|---|---|---|---|
| *Fukuoka March 2012* | | | | | | | | |
| OA (μg $m^{-3}$) | 1573 | 4.6 | 2.3 | 4.4 | 5.9 | 2.8 | 12.9 | LDL |
| m/z 44 (μg $m^{-3}$) | 1573 | 0.52 | 0.29 | 0.50 | 0.66 | 0.30 | 1.79 | LDL |
| LV-WSOC (μgC $m^{-3}$) | 11 | 1.670 | 0.763 | 1.516 | 2.092 | 1.116 | 3.034 | 0.689 |
| $\delta^{13}C$ of LV-WSOC (‰) | 11 | -25.0 | 2.9 | -25.0 | -24.0 | -26.4 | -19.7 | -30.3 |
| *Fukue March 2012* | | | | | | | | |
| OA (μg $m^{-3}$) | 1005 | 4.4 | 2.7 | 3.5 | 6.5 | 2.3 | 14.2 | 0.4 |
| m/z 44 (μg $m^{-3}$) | 1005 | 0.79 | 0.46 | 0.65 | 1.16 | 0.45 | 2.50 | 0.05 |
| LV-WSOC (μgC $m^{-3}$) | 11 | 2.390 | 1.370 | 1.645 | 3.466 | 1.387 | 5.139 | 0.900 |
| $\delta^{13}C$ of LV-WSOC (‰) | 11 | -21.9 | 1.2 | -21.8 | -21.1 | -22.8 | -20.2 | -23.7 |
| *Fukuoka December 2010* | | | | | | | | |
| OA (μg $m^{-3}$) | 1585 | 4.1 | 2.5 | 3.6 | 5.5 | 2.3 | 15.8 | LDL |
| m/z 44 (μg $m^{-3}$) | 1585 | 0.35 | 0.23 | 0.32 | 0.44 | 0.20 | 1.65 | LDL |
| LV-WSOC (μgC $m^{-3}$) | 8 | 0.686 | 0.280 | 0.666 | 0.829 | 0.512 | 1.138 | 0.273 |
| $\delta^{13}C$ of LV-WSOC (‰) | 8 | -23.2 | 1.7 | -23.3 | -22.3 | -24.2 | -20.6 | -26.0 |
| *Fukue December 2010* | | | | | | | | |
| OA (μg $m^{-3}$) | 1568 | 2.0 | 1.4 | 1.7 | 2.4 | 1.1 | 11.0 | LDL |
| m/z 44 (μg $m^{-3}$) | 1568 | 0.25 | 0.20 | 0.20 | 0.29 | 0.13 | 1.42 | LDL |
| LV-WSOC (μgC $m^{-3}$) | 11 | 1.082 | 0.464 | 1.029 | 1.102 | 0.837 | 2.110 | 0.401 |
| $\delta^{13}C$ of LV-WSOC (‰) | 11 | -22.0 | 2.0 | -21.2 | -20.5 | -23.4 | -19.3 | -25.4 |

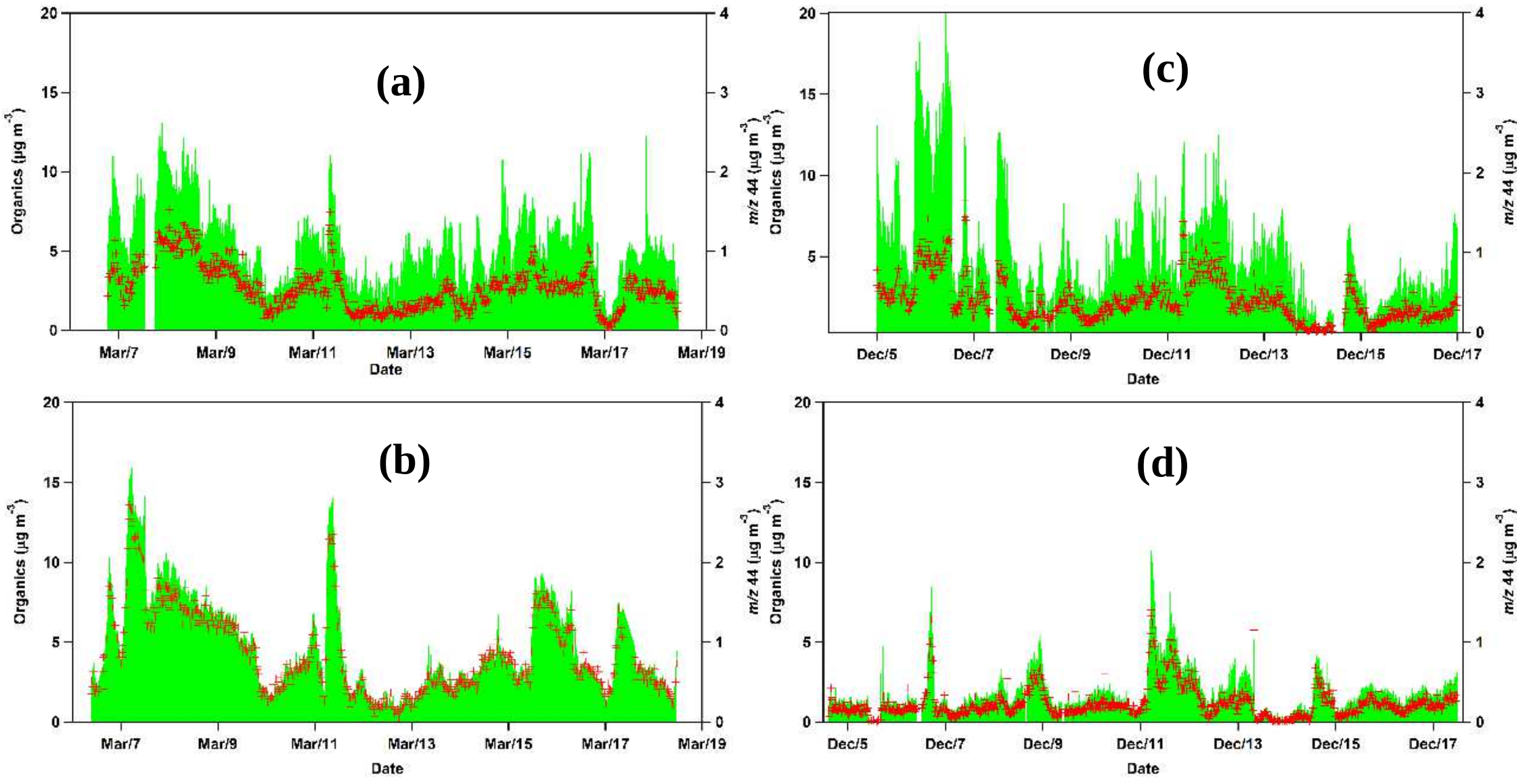

**Figure 3. Time series plot of OA (green filled) and m/z 44 (red cross) concentration observed by AMS measurements at (a) Fukuoka in March 2012, (b) Fukue in March 2012, (c) Fukuoka in December 2010, and (d) Fukue in December 2010.**

During the study period in December 2010, the level of OA concentrations at Fukuoka was a factor of two higher than the level at Fukue, while the level of *m/z* 44 concentration at Fukuoka was ~ 1.5 times higher than the level at Fukue (Table 1). Similarly to the results during the study in March 2012, the time series plots of OA and *m/z* 44 concentrations at Fukuoka showed similar variations to those at Fukue (Figure 3c and 3d) during the study, except for the period between December 5 and 8. Comparison of these plots with the time series plot of precipitation (Figure 2b) revealed that the precipitation did not always decrease those concentrations: the OA and *m/z* 44 concentrations at Fukuoka were high and those at Fukue were low during the period of December 5-8, even though more frequent precipitation was observed at Fukuoka during this period. These differences likely attribute to irregular contribution of primary OA from local emissions or SOA formed by oxidation of organics originating from local emissions at Fukuoka. To analyze the influence from local emission at Fukuoka in more detail, the observed concentrations at the two sites were directly compared.

The OA or *m/z* 44 concentrations at Fukuoka were plotted against those at Fukue to see if the observations at the two sites are correlated (Figure 4). In this scatter plot we compared the concentrations at Fukuoka that were acquired one or more hours behind the time that the concentrations at Fukue were acquired at. For the data sets obtained during the campaign in March 2012 and December 2010, the time lags used were one hour and two and half hours, respectively. The time lags were chosen because we have often observed pollution episodes in Fukuoka one or more hours later than the time the similar pollution episodes were observed in Fukue at. It was also because the plots with those time lags showed the highest coefficients of determination in their linear regression analysis. Plots for the OA and *m/z* 44 concentrations observed in March 2012 at Fukuoka and Fukue showed some degree of positive correlations between the two sites ($r^2$ of 0.37 and 0.47, respectively, shown in Figure 4a and 4b). Obtained linear regressions were $y = 0.45 \pm 0.02x + 2.6 \pm 0.1$ and $y = 0.36 \pm 0.01x + 0.23 \pm 0.01$ (± standard errors or SEs) for OA and *m/z* 44, respectively. The small, but significantly higher slope of OA than that of *m/z* 44 as well as the smaller $r^2$ value for OA than that for *m/z* 44 indicate different composition of OA between Fukuoka and Fukue, probably due to additional contribution of primary OA from local emissions to transboundary OA in the Fukuoka air. This speculation is supported by the results of PMF analysis discussed in the section 3.3 as well as the sharp rises and falls in the time series plot of OA concentrations (Figure 3a). However, one should be aware of that the dispersed plots can also be caused by different extent of atmospheric dilution, which depends on meteorological condition. The speculated contribution of local emission sources to the OA concentrations at Fukuoka should be considered as one of the factors dispersing the concentration plots. Nevertheless of the possible contribution of OA from local emissions to transboundary OA in the Fukuoka air, the first approximation shown in Figure 4a or 4b seems to be roughly capturing the overall trend, an indication of the roughly coinciding time-series variations of OA and *m/z* 44 concentrations between the two sites. Back trajectories of air masses modeled by HYSPLIT (Stein *et al.*, 2015) for each site exhibited the similar trajectories (Figure 5), indicating a possibility of long-range transport of air pollutants from the Asian continent. The proportional trends in Figure 4a and 4b together with the similar air mass trajectories at both sites suggest predominance of transboundary OA in the local air of Fukuoka. Meanwhile, there was almost no correlation in the OA and *m/z* 44 concentration plots between the two sites ($r^2$ of 0.049 and 0.219, respectively, shown in Figure 4c and 4d) during the study in December 2010, even though the HYSPLIT modeled the similar back trajectories of air masses from the Asian continent to both sites during the study period (Irei *et al.*, 2014, supporting information). Therefore, no valuable information was obtained from their linear regression analysis. Nevertheless of the insignificant correlations, it is worth noting that a part of OA and *m/z* 44 plots showed clear linear trends, slopes of which were substantially higher than one ($y = 11x$ for the OA and $y = 6x$ for the *m/z* 44 shown in Figure 4c and 4d, respectively). The data points of *m/z* 44 showing the relationship of $y = 6x$ were accompanied with the data points of OA showing the relationship of $y = 11x$, and those concentrations were clearly observed during the period of December 5-8. The synchronized increases suggest local episodes at Fukuoka, possibly associated with SOA.

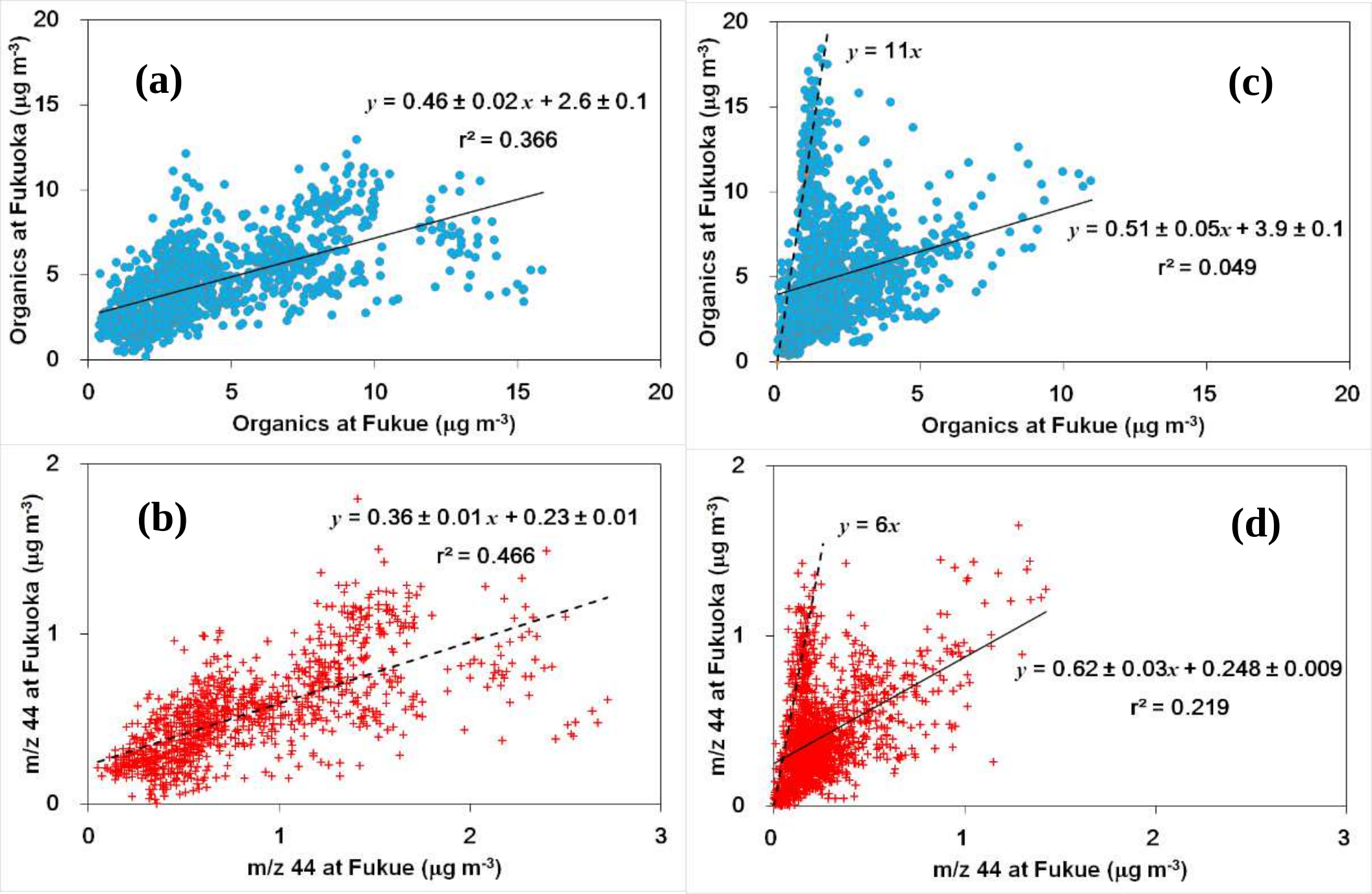

**Figure 4. Scatter plot of (a) OA and (b) m/z 44 concentrations observed at Fukue (x-axis) and Fukuoka (y-axis) during the field study in March 2012 and (c) OA and (d) m/z 44 concentrations during the field study in December 2010.**

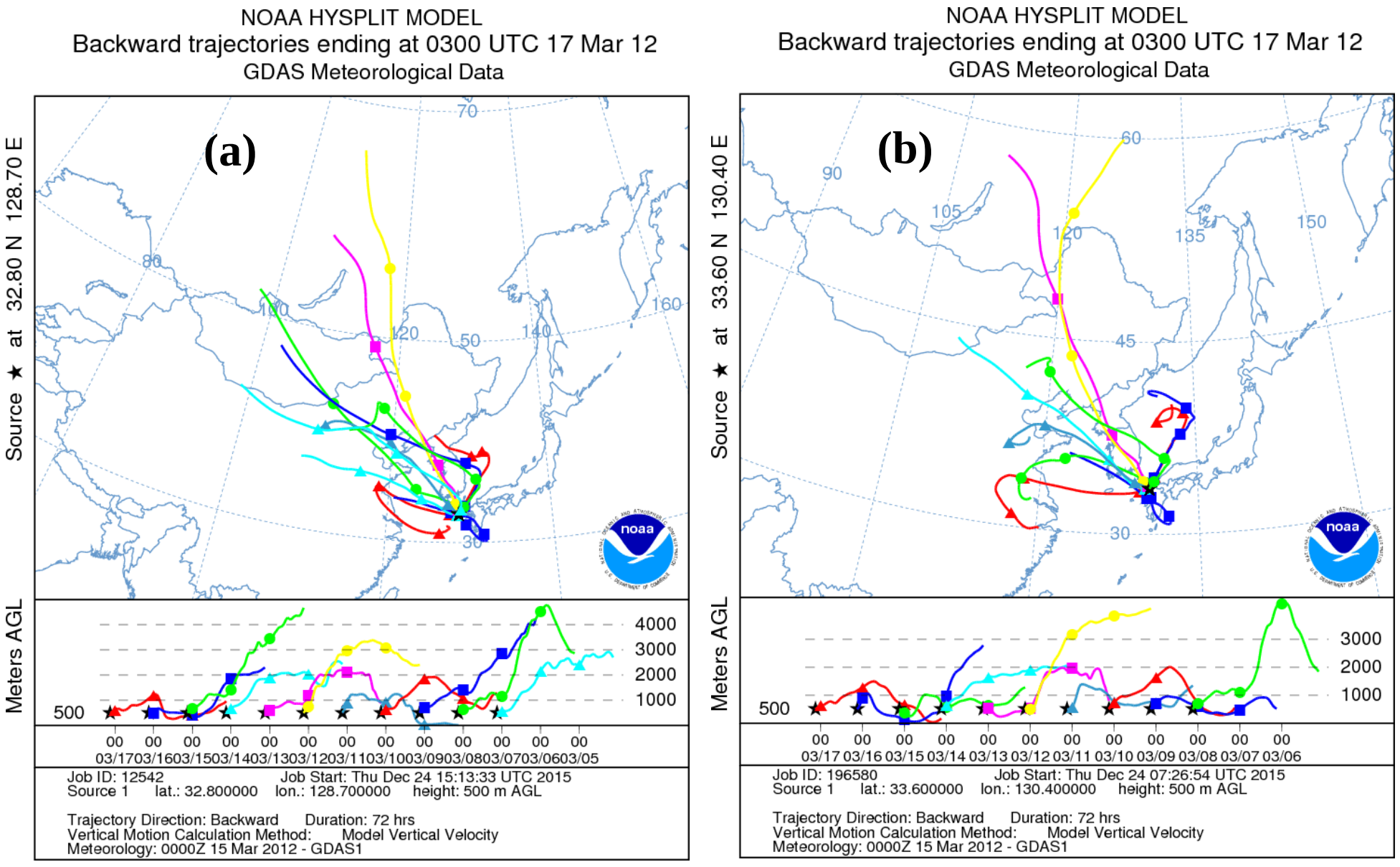

**Figure 5. Three-day back trajectories of air masses arriving at (a) Fukue and (b) Fukuoka City during the study period in March 2012.** The trajectories were drawn every 24 h.

Figure 6 shows time series plots of LV-WSOC concentrations and their $\delta^{13}C$ during the two study periods. The date of sampling start corresponds to the name of the filter sample. It should be noted that; the carbon concentration for the sample collected on March 13 2012 is not available due to a failure in the weighing procedure of the final extract that should have determined the extract volume; the measurement results for the samples collected on December 9th, 14th, and 15th 2010 are not available due to failures in their analysis. We should also note that all the carbon concentrations and $\delta^{13}C$ reported here were corrected for blank values using the blank carbon mass of 10 μgC per sample and its $\delta^{13}C$ of -17‰ for the Fukuoka data and 12 μgC per sample and -23 ‰ for the Fukue data. The concentration plots showed that, except for the sample collected on March 9, the LV-WSOC concentrations at Fukue were higher than those at Fukuoka through the two study periods. For the sample collected on March 9, detailed analysis on the LV-WSOC data discussed later implied that this sample may have had a problem, such as contamination. The higher concentrations observed at Fukue were more likely due to its proximity to the source(s) of LV-WSOC or its precursor(s): the closer to the sources, the higher concentration of airborne substances. Except for the samples collected on December 11 and 16 2010 and March 9 2012, the δ$^{13}$C values of LV-WSOC at Fukuoka were lower (*i.e.*, depleted in $^{13}$C) than those at Fukue. The lower δ$^{13}$C values at Fukue cannot be explained by a forward isotope fractionation of chemical reaction(s) of SOA precursor(s) in the gas-phase (Irei et al., 2006; Irei et al., 2011;

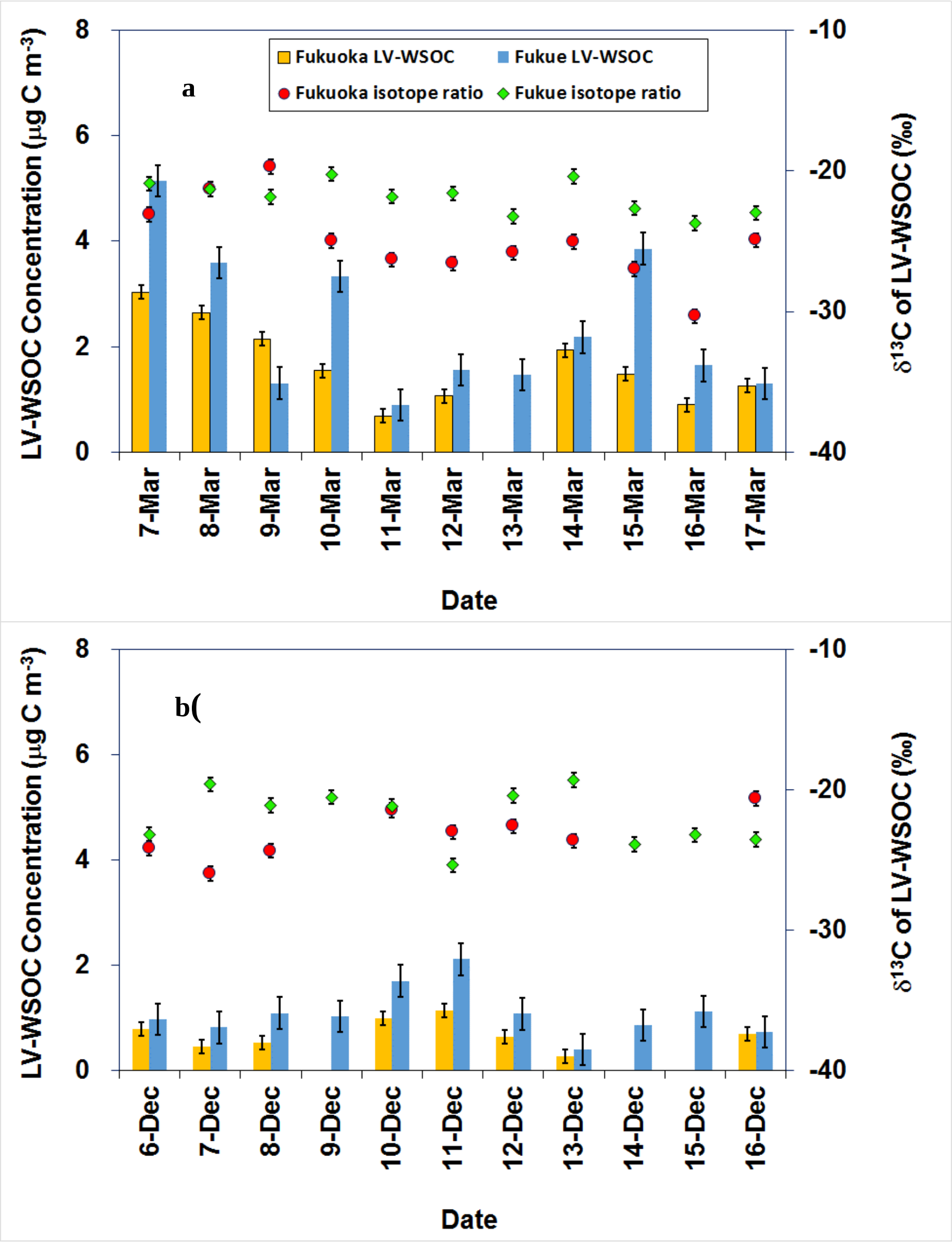

**Figure 6. Time series plot of the concentration and δ$^{13}$C of LV-WSOC observed at Fukuoka and Fukue during the field studies in (a) March 2012 and (b) December 2010.** The data points of December 9, 14, and 15 and March 13 at Fukue and of March are not available at Fukuoka.

Fisseha et al., 2009), neither by a forward isotope fractionation of chemical reaction(s) of organic substances in condensed phase (Betts and Buchannon, 1976; Pavuluri and Kawamura, 2012). Possible explanations for the lower $\delta^{13}C$ at Fukuoka are that; (1) there was LV-WSOC related to local emission sources with lower $\delta^{13}C$ values than that of the transboundary LV-WSOC or; (2) unknown process(s) with isotope fractionation(s) on transboundary LV-WSOC has shifted its $\delta^{13}C$ to lighter isotopic composition during the excess transporting time to Fukuoka.

Correlations were observed in the LV-WSOC concentrations between the two sites (Figure 7). The correlation was the highest ($r^2 = 0.832$) during the study in December 2010, regardless of the insignificant correlations of the OA and *m/z* 44 concentrations between the two sites previously discussed (Figure 4c and 4d). This discrepancy can be explained by semi-volatile organics originating from local emission in $PM_{1.0}$ at Fukuoka.

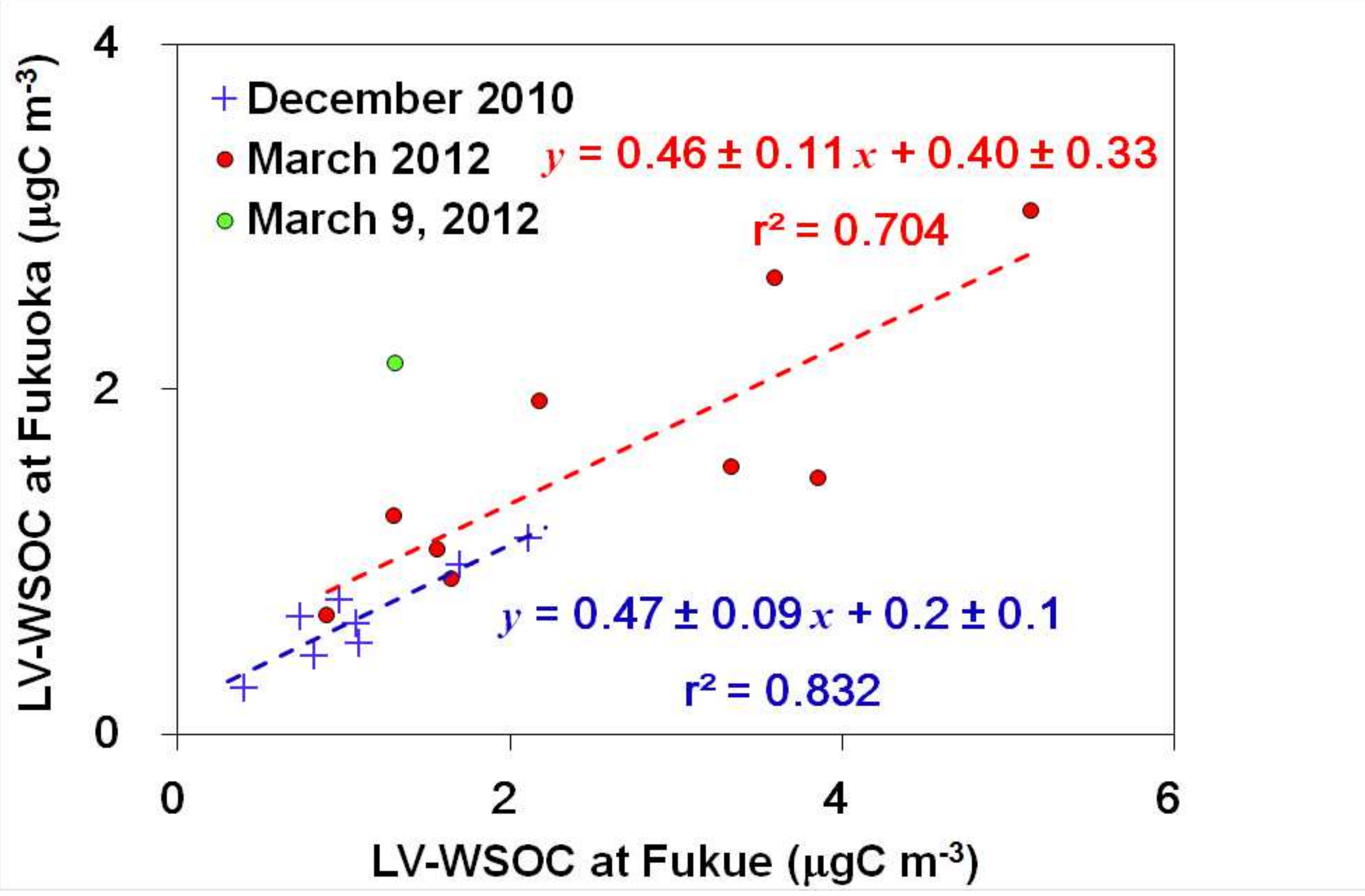

**Figure 7. Scatter plot of LV-WSOC concentrations observed at Fukue (x-axis) versus LV-WSOC concentrations observed at Fukuoka (y-axis) during the field study in March 2012 and December 2010.** The data point for the March 9 sample was excluded from the linear regression analysis.

This component was indicated by the results of PMF analysis discussed in the following section, and the component was evaporated out from LV-WSOC during the sample preparation process, resulted in consistency in the comparison plots of LV-WSOC between the two sites. Plot of LV-WSOC concentrations observed at two sites during the study in March 2012 also showed a positive correlation ($r^2$ = 0.517) when the data point for the March 9 sample was included and a higher correlation ($r^2$ = 0.704) when this data point was excluded. The high correlation between the two sites would indicate predominant influence of transboundary LV-WSOC in the Fukuoka air during the periods of these case studies.

### 3.3. PMF analysis of OA mass spectra

Using a PMF evaluation tool (Ulbrich *et al.*, 2009), we performed PMF analysis on the OA mass spectra obtained at Fukuoka during the study in March 2012. The following is a brief summary of PMF analysis for the studies previously reported. A two-factorial analysis gave the most feasible results for the dataset of December 2010 at Fukuoka and Fukue (Irei *et al.*, 2014), and these loadings were often identified as low-volatile oxygenated organic aerosol (LV-OOA with $f_{44}$ of 0.165) and hydrocarbon-like organic aerosol (HOA with $f_{44}$ of 0.002) (Figures 8a and 8b), which are known as aged oxygenated and unoxygenated OA, respectively. The mass spectra for HOA are very similar to those for semi-volatile oxygenated organic aerosol (SV-OOA), known as fresh oxygenated OA. The difference can be found in SV-OOA mass spectra if there is a mass spectrum with small abundance at *m/z* 44. According to this feature, the HOA loading at Fukuoka shown in Figure 8a could be SV-OOA, which was found sometime in the Fukuoka and Fukue air (Takami *et al.*, 2016; Irei *et al.*, 2016). During the study in March 2012 at Fukue, one-factorial analysis resulted in the most reasonable solution (Irei *et al.*, 2015), with the extracted loading identified as LV-OOA (Figure 8c).

PMF analysis on the OA mass spectra observed during the field study in March 2012 at Fukuoka showed that a two-factorial analysis gave the most reasonable results: loadings corresponding to LV-OOA and HOA (Figure 8d), with average fractions of 72% and 28% and $f_{44}$ of 0.165 and 0.002, respectively. In this PMF analysis, the signal at *m/z* 27 was excluded due to the large amount of noise caused by $N_2$ at the edge of the slightly wider peak width at *m/z* 28. A comparison of the deconvoluted loadings for the Fukuoka data with the previously reported single loading at Fukue during the study in March 2012 revealed that the LV-OOA reference mass spectra from the AMS mass spectra database by Ulbrich et al. (http://cires.colorado.edu/jimenez-group/AMSsd/) and the LV-OOA loading at Fukue during the study in March 2012 (Irei et al., 2015) coincided with the LV-OOA loading at Fukuoka in this study ($r^2$ = 0.962 and 0.967, respectively). The HOA reference mass spectra and the HOA loading from the previous study also coincided with the HOA loading at Fukuoka in this study ($r^2$ = 0.576 and 0.805, respectively).We concluded that HOA at Fukuoka was local origin because there was no such a loading deconvoluted from the OA mass spectra obtained at Fukue during this study period (Irei et al., 2015).

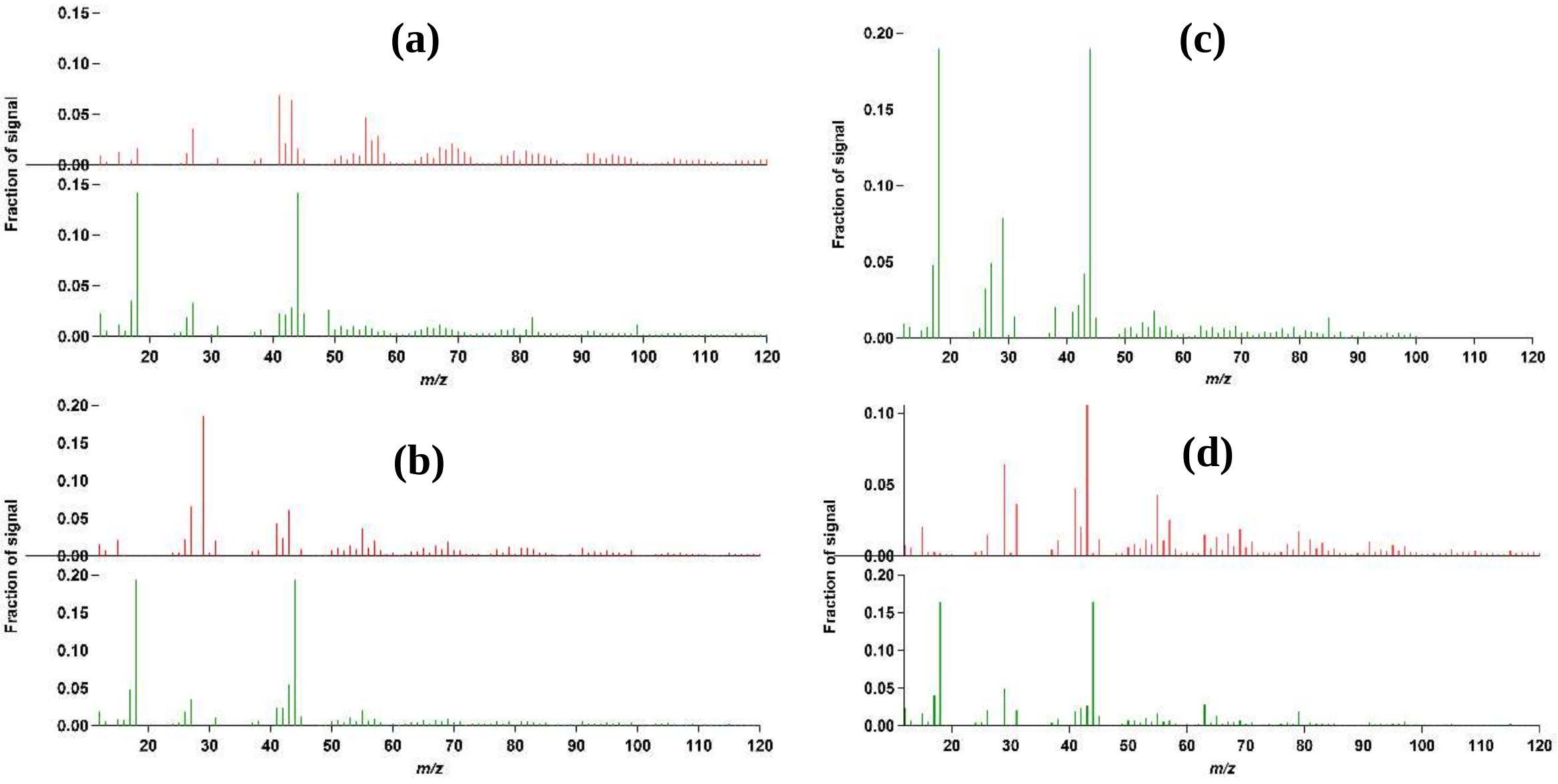

**Figure 8. Loading (m/z 12 to m/z 120) yielded from PMF analysis on organic mass spectra for (a) the two factorial solution of Fukuoka December, 2010; (b) two factorial solution of Fukue December, 2010; (c) one factorial solution of Fukue March, 2012; and (d) two factorial solution of Fukuoka March, 2012.** The extracted loadings shown are LV-OOA (green) and HOA or SV-OOA (red).

Using the method used by Zhang *et al.* (2005), we estimated ratios of organic mass (OM) to organic carbon (OC) for each of the deconvoluted loadings produced by the PMF. The OM / OC ratios for LV-OOA and HOA during the study in March 2012 at Fukuoka were 4.7 and 2.9 µg µg$C^{-1}$, respectively. Compared to the OM / OC ratios for LV-OOA and HOA previously observed at Fukue and Fukuoka (Table 2), these values are the closest to the average OM / OC ratios for LV-OOA and HOA during the half-year study period at Fukue. The ratio for LV-OOA also corresponded approximately to the ratio of 4.3 µg µg$C^{-1}$ for LV-OOA observed at Fukue. Using these OM / OC ratios and the fractions of LV-OOA and HOA determined above, the overall OM / OC ratio for OA (*i.e.*, the mixture of LV-OOA and HOA) was calculated to be 4.2 µg µg$C^{-1}$. This overall ratio was used later for a comparison with an OA / LV-WSOC ratio, which will be discussed in the following section.

**Table 2. OM/OC ratio calculated for PMF loadings**

| Field studies | LV-OOA | HOA |
|---|---|---|
| December 2010, Fukue[b] | 3.6 | 1.2 |
| December 2010, Fukuoka[b] | 3.5 | 1.2 |
| December 2010, Hedo[b] | 3.8 | n.a.[a] |
| | | |
| March 2012, Fukue[c] | 4.3 | n.a.[a] |
| March 2012, Fukuoka | 4.7 | 2.9 |
| | | |
| December 2010 through May 2011, Fukue[d] | 5.0 | 2.8 |

[a] Not available because the PMF analysis ended with only the LV-OOA loading.
[b]Adopted from Irei et al. (2014). [c]Adopted from Irei et al. (2015). [d]Adopted from Irei et al. (2016).

### 3.4. Comparison of LV-WSOC with OA and *m/z* 44

The LV-WSOC concentrations were compared with the *m/z* 44 concentrations to have an idea of the composition of LV-WSOC. It was found that the LV-WSOC determined by the filter sample analysis and the *m/z* 44 concentrations determined by the AMS measurements during the study in March 2012 were highly correlated at Fukue ($r^2$ = 0.91 with a slope of 0.28 ± 0.03 μg μgC$^{-1}$) and Fukuoka ($r^2$ = 0.75 with a slope of 0.30 ± 0.06 μg μgC$^{-1}$) (Figures 9a and 9b). It was also found that the LV-WSOC and OA concentrations in this study were also highly correlated at Fukue ($r^2$ = 0.94 with a slope of 1.65 ± 0.14 μg μgC$^{-1}$) and marginally correlated at Fukuoka ($r^2$ = 0.60 with a slope of 2.02 ± 0.58 μg μgC$^{-1}$). For the plot of OA (and *m/z* 44) versus LV-WSOC at Fukuoka, one data point (the outlying LV-WSOC concentration of March 9 shown in Figure 6a) significantly lowered the $r^2$ values. There was no AMS measurement problem during a period that determined the *m/z* 44 and OA concentrations for March 9. Excluding this data point, the slope and $r^2$ were improved to 0.33 ± 0.03 μg μgC$^{-1}$ and 0.92 for *m/z* 44, and 2.3 ± 0.3 μg μgC$^{-1}$ and 0.88 for OA. This improvement indicates inconsistency of the filter sample collected on March 9, and it likely implies that the sample may have had something problem, likewise contamination during the sample handling procedure.
Even though a possibility of LV-WSOC contribution from primary emission sources to this sample at Fukuoka cannot be excluded, there was no such indication in the results of the AMS measurements and of the PMF analysis. Linear regressions in Figure 9a and 9b exhibited the insignificantly different slopes for the plots of *m/z* 44 versus LV-WSOC at Fukuoka and Fukue, suggesting that *m/z* 44 and LV-WSOC at both sites are likely from the same origin, probably transboundary SOA. Meanwhile, the slope of the linear regressions for the plots of OA versus LV-WSOC at Fukuoka (2.3 ± 0.3 μg μgC$^{-1}$) was slightly, but significantly higher than that at Fukue (1.7 ± 0.1 μg μgC$^{-1}$), implying some additional contribution of OA from local emission sources at Fukuoka. For the field study in December, 2010, the LV-WSOC concentrations were also highly correlated with the *m/z* 44 concentrations at Fukuoka ($r^2$ = 0.88 with a slope of 0.58 ± 0.10 μg μgC$^{-1}$) and Fukue ($r^2$ = 0.85 with a slope of 0.22 ± 0.03 μg μgC$^{-1}$), as well as with the OA concentrations at Fukuoka ($r^2$ = 0.68 with a slope of 4.4 ± 1.2 μg μgC$^{-1}$) and Fukue ($r^2$ = 0.83 with a slope of 1.45 ± 0.22 μg μgC$^{-1}$) (Irei *et al.*, 2014). The slope for the regression of OA versus LV-WSOC at Fukuoka was significantly higher than that at Fukue in this study. Unlike the slopes obtained during the study in March 2012, the slope for the regression of *m/z* 44 versus LV-WSOC was also significantly higher. The higher slope was caused by the higher *m/z* 44 concentrations with the lower LV-WSOC concentrations at Fukuoka. A plausible explanation would be that this significantly higher slope during the study in December 2010 at Fukuoka was due to the additional contribution of *m/z* 44 from local origin, possibly SV-OOA, to transboundary *m/z* 44, but not to LV-WSOC. The OA / LV-WSOC ratio in March 2012, one of the linear regression slopes shown in Figure 9a, was further compared with the previously discussed OM / OC ratio of OA at Fukuoka to know the differences in OC concentration between TSP and $PM_{1.0}$.

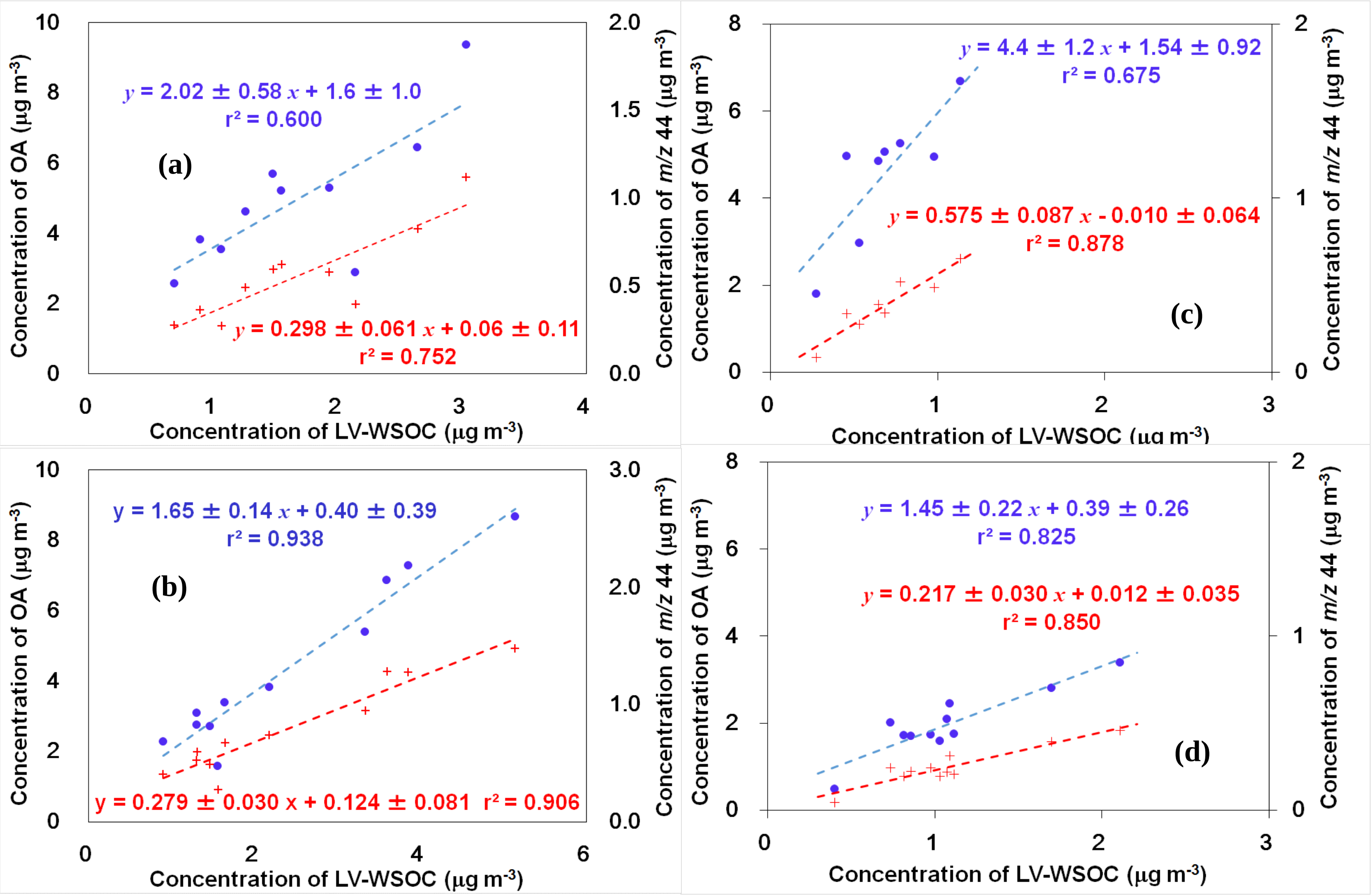

**Figure 9. Plot of LV-WSOC concentration from the filter sample analysis versus the OA concentration (blue) and m/z 44 concentration (red) measured by AMS at (a) Fukuoka and (b) Fukue in March 2012 and (c) Fukuoka and (d) Fukue in December 2010.**

The OA / LV-WSOC ratio of 2.0 μg $\mu gC^{-1}$ for the Fukuoka data was compared with the previously determined overall OM / OC ratio of 4.2 μg $\mu gC^{-1}$ to better understand the difference between OC concentrations in TSP and in $PM_{1.0}$ measured by the AMS. The OA / LV-WSOC ratio was found to be a half of the OM / OC ratio, implying that the TSP contained more than a factor of two as much OC as $PM_{1.0}$ contained. Nevertheless of missing such substantial quantity of OC that are compared with LV-WSOC, the high correlations between the LV-WSOC in TSP and the OA or *m/z* 44 in $PM_{1.0}$ suggest that OC was distributed widely over the various range of particle size. In other words, the LV-WSOC observed was likely SOA partitioned to pre-existing aerosols distributing in a wide range of particle size. Further evidence for the predominance of SOA was found in the $\delta^{13}C$ of LV-WSOC.

Plots of the $\delta^{13}C$ values of LV-WSOC versus the $f_{44}$ of OA showed a positive correlation ($r^2 = 0.646$) (Figure 10). The systematic increase of $\delta^{13}C$ values with various $f_{44}$ (i.e., various extent of oxidation processing in this case) is the evidence of SOA formation found by laboratory studies (Irei et al., 2006; Irei et al., 2011). Such a systematic trend was observed first in ambient measurements for $\delta^{13}C$ of LV-WSOC at Cape Hedo in Okinawa (Irei et al., 2014), and this study is the second case. It is known that the systematic increase of $\delta^{13}C$ with the $f_{44}$ is known to be case-dependent (Irei *et al.*, 2015). In the report they proposed that $f_{44}$ could be used as an oxidation indicator qualitatively under the condition where LV-OOA with a high $f_{44}$ (*e.g.*, an SOA origin) was mixed with a constant amount of HOA that has a low $f_{44}$. This hypothesis fully supports our observations in Figure 10 because the PMF analysis discussed in section 3.3 reasonably deconvoluted the OA into LV-OOA and HOA. With consideration of the strong association of *m/z* 44 with LV-WSOC, the similar *m/z* 44 ratios to LV-WSOC at Fukue and Fukuoka, and the similar back trajectories of air masses from the Asian continent to Fukue and Fukuoka, we conclude that the majority of LV-WSOC observed at Fukuoka during the study in March 2012 was made of transboundary SOA. As discussed in previous publication (Irei et al., 2014), we did not observed such a systematic increase in the plot of $\delta^{13}$C for the study in December 2010. Even though it is possible that the LV-WSOC in that study was also made of transboundary SOA, the evidence in the plot of $\delta^{13}$C was likely vailed with SV-OOA.

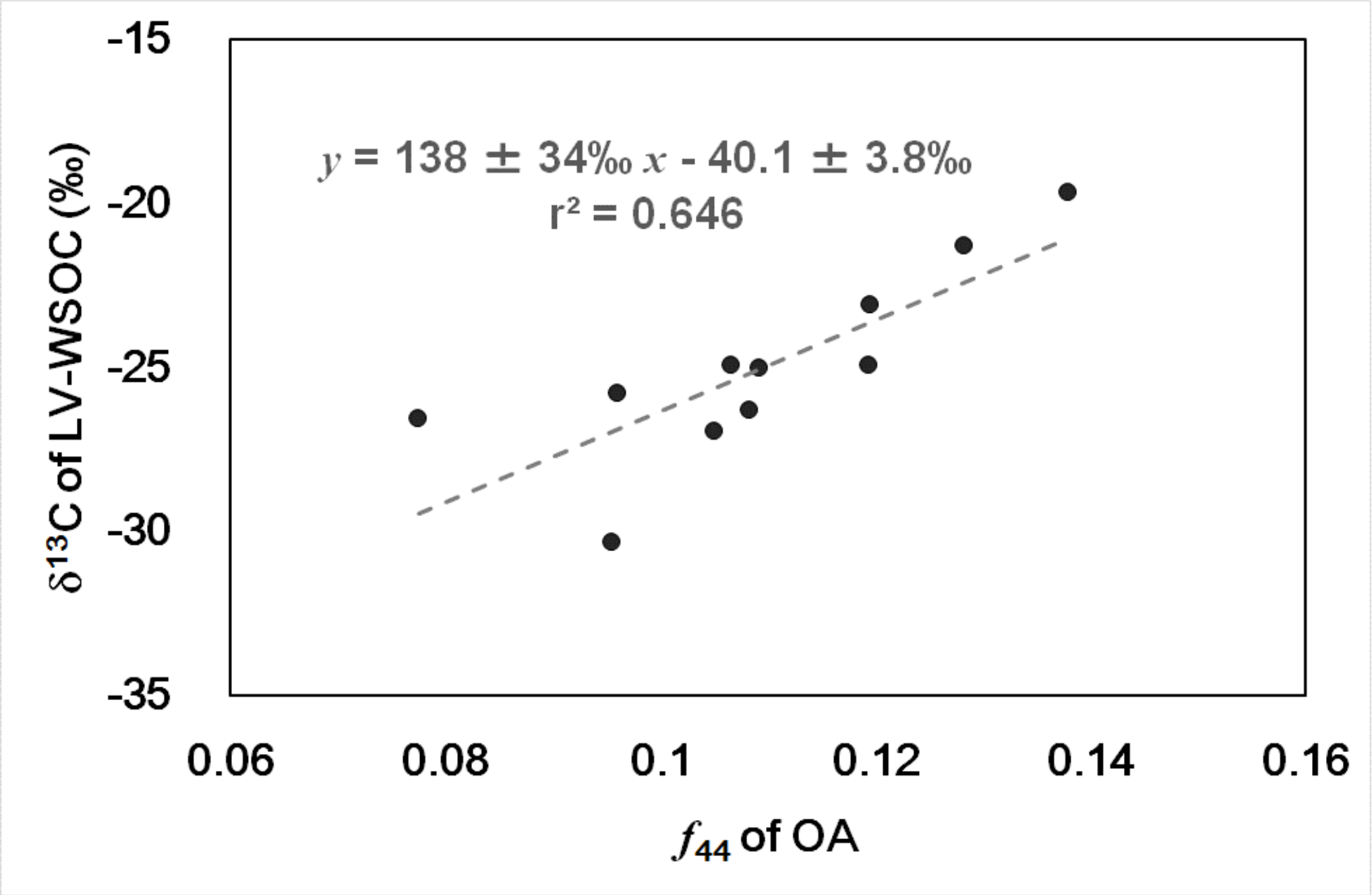

**Figure 10. Plot of $\delta^{13}$C of LV-WSOC from the filter sample analysis versus $f_{44}$ of organic aerosol measured by AMS at Fukuoka, March 2012.**

The linear regression in Figure 10 was $\delta^{13}C = (138 \pm 34)‰ \times f_{44} - (40.1 \pm 3.8)‰$. Compared to the linear regression previously observed, the slope here was more than five times higher. The magnitude of slope depends on kinetic isotope effects (KIEs) of chemical reactions during the loss and production of particulate carboxylic acids. This five times higher slope possibly suggest the production of SOA via oxidation of different precursor(s) from those in the previous studies. The intercept, -40.1‰ ± 3.8‰, may be the $\delta^{13}$C of background LV-WSOC or the $\delta^{13}$C of SOA produced at a very early stage of precursor oxidation. More detailed studies will be needed to understand what this background value indicates.

When a targeted organic species observed at an urban site originates from multiple emission sources (*i.e.*, a case of two or more member mixing), one of which is a common source for the substance observed at a background site, a quotient of its concentrations at two sites can be used as a scale for evaluation of predominant contribution of transboundary SOA to the urban air quality in Fukuoka. We tested this hypothesis using *m/z* 44 ratios: the LV-WSOC and OA concentrations were plotted as a function of ratio of daily averaged *m/z* 44 concentration at Fukuoka to that at Fukue (Figure 11). Similarly to the section 3.2, time-lagged *m/z* 44 and OA concentrations were used here to calculate daily averaged concentrations of *m/z* 44 and OA at Fukuoka for the comparison with the *m/z* 44 data at Fukue. In this plot, a *m/z* 44 ratio smaller than one should indicate predominance of carboxylate from mid- or long-range transport origin, and a *m/z* 44 ratio higher than one should indicate predominant contribution of carboxylate from other source(s). The plot exhibits high *m/z* 44 and LV-WSCO concentrations as the *m/z* 44 ratios were below one. In contrast, those concentrations were low as the *m/z* 44 ratios were higher than one. This means that even in the urban air of Fukuoka the influence from long-range transported SOA on the concentrations of particulate carboxylate in $PM_{1.0}$ and LV-WSOC in TSP was large during the two study periods. Meanwhile, there was no dependency between the concentrations of OA and the *m/z* 44 ratios. This is more likely explained by a substantial contribution of SV-OOA originating from local emission source(s) at Fukuoka, which was not contained in LV-WSOC.

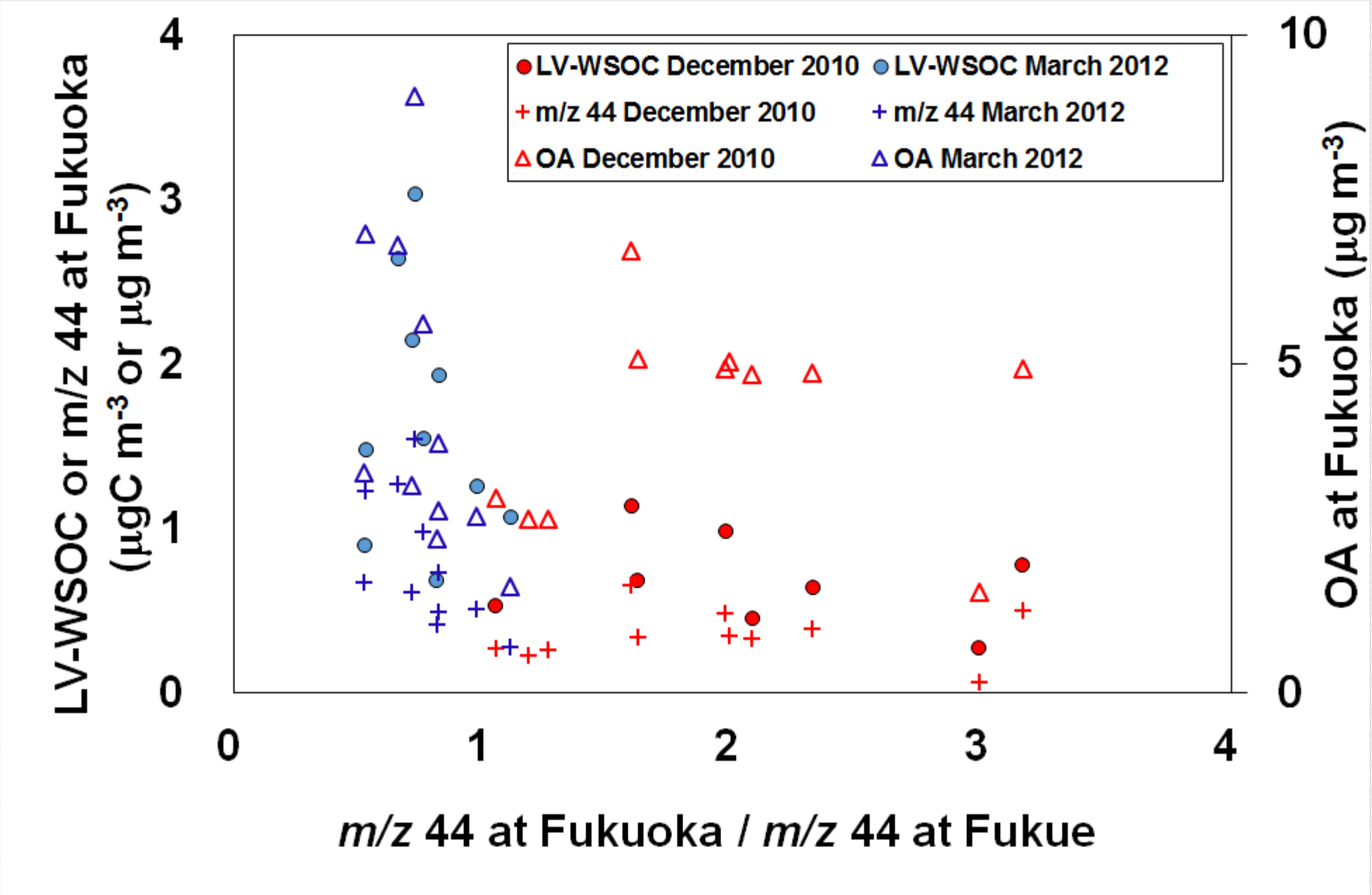

**Figure 11. Plot of daily average concentration of m/z 44, OA, and LV-WSOC as a function of ratio of daily average *m/z* 44 concentration at Fukuoka to daily average *m/z* 44 concentration at Fukue.**

**Summary**

To evaluate influence of transboundary SOA on the urban air of Fukuoka, we conducted simultaneous field studies in December 2010 and March 2012 on Fukue Island and in the city of Fukuoka, representing a rural and an urban sites of northern Kyusyu region. In the studies, OA in $PM_{1.0}$ was analyzed by AMSs, and obtained mass spectra for OA were analyzed by a PMF method to identify and quantify OA components. Independently, airborne TSP was collected on filters, and LV-WSOC extracted from the TSP filter samples were analyzed by an EA-IRMS for its concentration and $\delta^{13}C$. The PMF analysis resulted in that the OA at Fukuoka contained predominant amount of transboundary LV-OOA and marginal amount of HOA during the study in March 2012 and LV-OOA and SV-OOA in the same magnitude during the study in December 2010. The comparison of LV-WSOC concentrations in TSP with the *m/z* 44 concentrations in $PM_{1.0}$ showed a high correlation ($r^2 > 0.91$) with a slope of ~ 0.3 μg μgC$^{-1}$ through the study in March 2012 at both sites, indicating predominant contribution of transboundary LV-WSOC to the LV-WSOC in the Fukuoka air. Furthermore, the plot of $\delta^{13}C$ of LV-WSOC as a function of extent of oxidation processing, $f_{44}$, showed a systematic increasing trend, an indication of SOA. The series of findings suggest that the observed LV-WSOC and *m/z* 44 originated from transboundary SOA and was the predominant component of LV-WSOC in the Fukuoka air during the case study in March 2010. Meanwhile, the measurement results during the study in December 2010 showed different slopes for the linear regressions for the Fukuoka and Fukue data drawn from the plots of LV-WSOC versus *m/z* 44 (0.58 and 0.22 μg μgC$^{-1}$, respectively), with high correlations ($r^2 > 0.85$). With consideration of the results from PMF analysis, we concluded that there was likely a contribution of SV-OOA to transboundary *m/z* 44, but not to LV-OOA. This contribution made $f_{44}$ not to work as an oxidation indicator during this study, resulting in the fact that the contribution of transboundary SOA to the LV-WSOC in the Fukuoka air could not be evaluated. More extensive studies are needed to understand the transboundary SOA in the Fukuoka air in depth.

**Acknowledgements**

We acknowledge the NOAA Air Resources Laboratory (ARL) for the provision of the HYSPLIT transport and dispersion model and/or READY website (http://www.ready.noaa.gov) used in this publication. This project was financially supported by the Internal Encouraging Research Fund for Early Career Scientists at the Center for Regional Environmental Research in the National Institute for Environmental Studies. The project was also partially supported by a Grant-in-Aid for Scientific Research on Innovative Areas (No. 4003) from the Ministry of Education, Culture, Sports, Science and Technology, Japan, the International Research Hub Project for Climate Change and Coral Reef/Island Dynamics of Univ. of the Ryukyus, and the ESPEC Foundation for Global Environment Research and Technologies (Charitable Trust).